\documentstyle[11pt,paspconf,epsf]{article}
\begin{document}

\title{CLUSTER IDENTIFICATION VIA VORONOI TESSELLATION}

\author{M. Ramella\altaffilmark{1}, M. Nonino, W. Boschin}
\affil{Osservatorio Astronomico di Trieste, via Tiepolo 11, 34131
Trieste, Italy} 
\author{D. Fadda\altaffilmark{2}} 
\affil{Service d'Astrophysique, CEA Saclay-Orme de Merisiers,
F91191 Gif-sur-Yvette Cedex, France} 
\altaffiltext{1}{e.mail:ramella@oat.ts.astro.it} 
\altaffiltext{2}{e.mail:fadda@hubble.saclay.cea.fr}

\begin{abstract}
We propose an automated method for detecting galaxy clusters in
imaging surveys based on the Voronoi tessellation technique. It
appears very promising, expecially for its capability of detecting
clusters indipendently from their shape. After a brief explanation of
our use of the algorithm, we show here an example of application based
on a thin strip centered on the ESP Key Programme complemented with
galaxies of the COSMOS/UKST Southern Sky Catalogue.
\end{abstract}


\keywords{galaxy clusters, cosmology}

\section{Introduction}

Wide field imaging is becoming increasingly common since new large
format CCD cameras are (or soon will be) available at several
telescopes. In particular, wide field imaging of the extragalactic sky
allows systematic searches for clusters of galaxies. Reliable 2D
catalog of clusters are of great cosmological importance since they
are starting points for the study of the largest virialized density
fluctuations in the universe.  Several algorithms have been developed
so far for the 2D identification of clusters, for example wavelets
(Fadda et al. 1997), matched filters (Postman et al. 1996), adaptive
smoothing (Pisani 1996).  We propose a method for the identification
of clusters based on the Voronoi tessellation. The main advantages of
our method are the following: a) it is fast, b) it is completely
non-parametric, c) galaxies are naturally assigned to structures.  In
section 2 we briefly summarize the Voronoi algorithm, while in section
3 we present an example of application of the method on galaxies in a
thin strip centered on the ESP Key Programme (Vettolani et al. 1997)
complemented with galaxies of the COSMOS/UKST Southern Sky Catalogue 
supplied by the Anglo-Australian Observatory.

\section{The algorithm}

The Voronoi tessellation creates a partition of the plane based on a
2-D distribution of points. The algorithm assignes to each point the
region of the plane whose points are nearer to this point than to any
other point (Figure 1). After having associated Voronoi tessels to all
points, we compute a distribution of densities (see Figure 2, solid
line). We define as density the inverse of the area of a Voronoi
tessel:
\begin{equation}
density = 1 / Voronoi\  area
\end{equation}

\begin{figure}
\centering
\leavevmode
\epsfxsize=.75\textwidth
\epsfbox{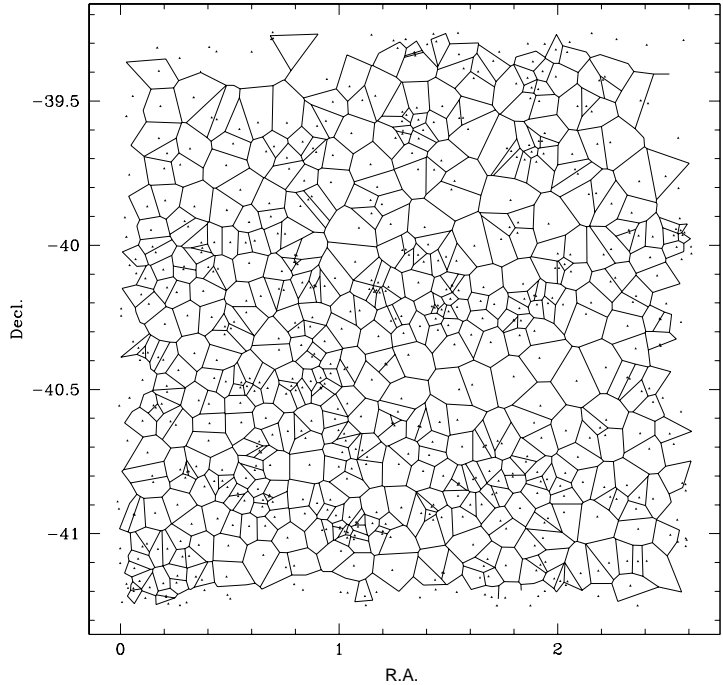}
\caption{Partition of a plane by the Voronoi tessellation algorithm.}
\end{figure}
Clearly, clustered points are high density regions while ``background''
points constitute the low-density tail of the density distribution. 
Before
proceeding further, we reject all boundary-points because they have an
infinite area.
Then, following Ebeling (1993), we fit the low-density
tail of our observed distribution with the analytical form of the
density distribution of a poissonian distribution of points (Kiang
1966). We assume that the fitted Kiang distribution
(Figure 2, dashed line) is the density distribution of the
background points and derive
the density threshold for the detection of ``clusters'' at a chosen 
significance level, usually 95\% and 99\% (Figure 2, vertical lines). 
A cluster output by our procedure consists of an ensamble of data-points
with their tessels (cluster members). 
Before accepting the cluster in the final catalog
we add to the list of cluster members a number of points that 
produce a more regularly shaped cluster. For example, we list as members 
all points within the ``convex hull'' of the original cluster.
Finally, we fit an ellipse to the cluster with the requirements that
the area of the ellipse equals the total tessel area of the cluster members
(Figure 3).
\begin{figure}
\centering
\leavevmode
\epsfxsize=.68\textwidth
\epsfbox{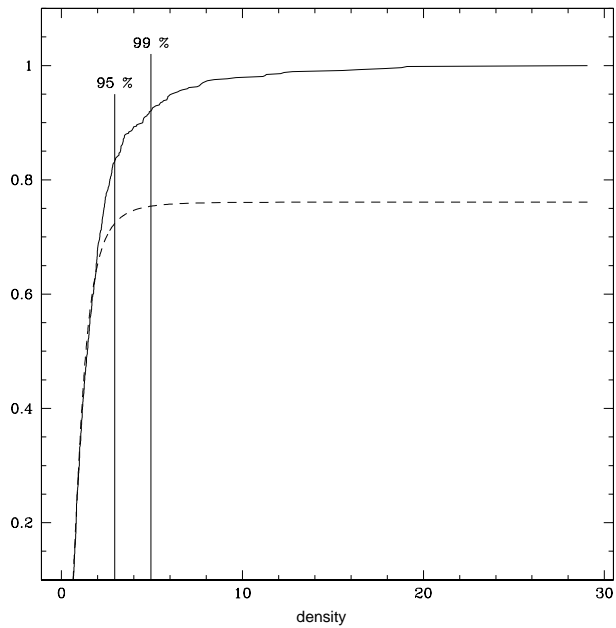}
\caption{The distribution of densities in the plane drawn in Fig.1 (solid line) and the fitted Kiang distribution (dashed line). Two density thresholds for the detection of clusters are drawn too.}
\end{figure}

\section{An example}

As a first test of our method, we search for clusters in a thin strip
overlapping the ESP Key-Programme (Vettolani et al. 1997). We
complement the ESP data with galaxies of the COSMOS/UKST Southern Sky
Catalogue supplied by the Anglo-Australian Observatory. The
limits of the strip are $23^{h} 23^m \le \alpha_{1950} \le 01^{h} 20^m
$ and $22^{h} 30^m \le \alpha_{1950} \le 22^{h} 52^m $ respectively; $
-40^o 45' \le \delta_{1950} \le -39^o 45'$. We consider only galaxies
brighter than b$_j = 19.4$. In this strip we find 18 fluctuations at
the 99\% confidence level. Nine of these fluctuations (50\%)
correspond to clusters listed in the EDCC (Lumsden et al. 1992) with
centers in the ESP strip. The only EDCC cluster we do not identify is
EDCC 185 (=ACO supplementary list 1055) becasue of its very low
redshift z = 0.0322 and the narrow width of our strip.
\begin{figure}
\centering
\leavevmode
\epsfxsize=0.65\textwidth
\epsfbox{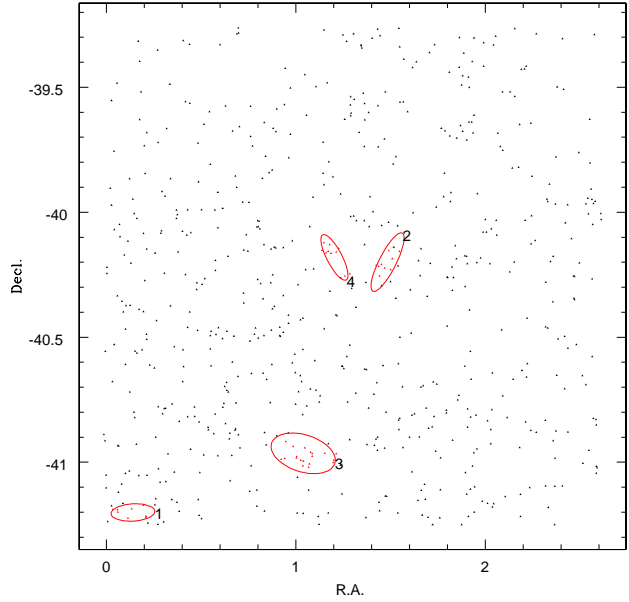}
\caption{The ellipses corresponding to the clusters found by the Voronoi algorithm at the 95 \% c.l.}
\end{figure}
\begin{figure}
\centering
\leavevmode
\epsfxsize=.65\textwidth
\epsfbox{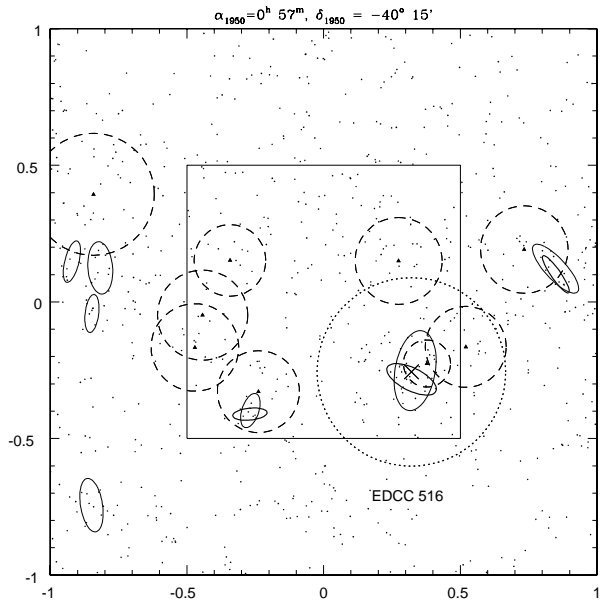}
\caption{Identified clusters on a 2 degrees window centered on a 1 degree field
in the ESP strip. Ellipses are our clusters at 95 \% c.l. (solid) and 98 \% c.l. (thick solid), dotted circles are EDCC clusters and dashed circles are ESP groups. Note the cluster EDCC 516.}
\end{figure}
Six fluctuations (0.33\%) correspond to rich groups of galaxies
objectively identified in redshift-space (Ramella et al. 1998).  These
groups have at least 5 members and are more distant than z=0.05.
Three structures (17 \%) have no counterpart in the EDCC or ESP group
catalogs. Figure 4 is a 2 degrees window centered on a 1 degree field
in the ESP strip. Dots are galaxies, ellipses are our clusters,
crosses at the center of dotted circles are EDCC clusters, the
triangles at the center of dashed circles are ESP rich groups.  We are
currently running a more refined version of the algorithm on much
deeper fields and on simulations of distant clusters.


\begin{references}
\reference Ebeling, H., 1993, Ph.D. Thesis 
\reference Fadda, D., Slezak, E. \& Bijaoui, A., 1997, \astap, 127, 335
\reference Kiang, T., 1966, Z.f. Astroph. 64, 433
\reference Lumsden, S.L., Nichol, R.C., Collins, C.A. \& Guzzo, L., 1992, \mnras, 258, 1
\reference Pisani, A., 1996, \mnras, 278, 697
\reference Postman, M., Lubin, L., Gunn, J.E., Oke, J.B., Hoessel, J.G., Schneider, D.P. \& Christensen, J.A., 1996, \aj, 111, 615
\reference Ramella, M., et al., 1998, \astap, in press
\reference Vettolani, G., et al., 1997, \astap, 325, 954
\end{references}
\end{document}